\documentclass[useAMS,usenatbib]{mn2e}
\usepackage{graphicx}

\footnotesize
\newdimen\digitwidth    
\setbox0=\hbox{\rm0}
\digitwidth=\wd0
\catcode`!=\active
\def!{\kern\digitwidth}
\normalsize

\newcommand{\axp}{AXP\,J1810$-$197\,}

\title{Radio spectrum of the AXP~J1810$-$197 and of its profile components}
\author[K.Lazaridis et al.]
{
K.~Lazaridis,$^1$
 A.~Jessner,$^1$ M.~Kramer,$^2$ B.~W.~Stappers,$^2$$^,$$^3$ A.~G.~Lyne,$^2$ 
\newauthor \ C.~A.~Jordan,$^2$ M.~Serylak$^3$$^,$$^4$ and J.~A.~Zensus$^1$
\\
$^{1}$Max-Planck-Institut f\"ur Radioastronomie, Auf dem H\"ugel 69, 53121, Bonn, Germany
\\
$^{2}$University of Manchester,
Jodrell Bank Centre for Astrophysics, Alan-Turing Building, Manchester M13
9PL, UK
\\
$^{3}$Stichting ASTRON, Postbus 2, 7990 AA, Dwingeloo, the Netherlands
\\
$^{4}$Astronomical Institute "Anton Pannekoek", University of Amsterdam,
Kruislaan 403, 1098 SJ Amsterdam, The Netherlands
}
\let\leqslant=\leq 
\date{}
\begin{document}

\maketitle
\newcommand{\setthebls}{
}

\setthebls

\begin{abstract} 
  As part of a European Pulsar Network (EPN) multi-telescope observing
  campaign, we performed simultaneous multi-frequency observations at
  1.4, 4.9 and 8.4 GHz during July 2006 and quasi-simultaneous
  multi-frequency observations from December 2006 until July 2007 at
  2.7, 4.9, 8.4, 14.6 and 32 GHz, in order to obtain flux density
  measurements and spectral features of the 5.5-sec radio-emitting
  magnetar \axp. We monitored the spectral evolution of its pulse
  shape which consists of a main pulse (MP) and an interpulse (IP). We
  present the flux density spectrum of the average profile and of the
  separate pulse components of this first-known radio-emitting
  transient anomalous X-ray pulsar.  We observe a decrease of the flux
  density by a factor of 10 within 8 months and follow the
  disappearance of one of the two main components. Although the
  spectrum is generally flat, we observe large fluctuations of the
  spectral index with time. For that reason we have made some measurements
  of modulation indices for individual pulses in order to also
  investigate the origin of these fluctuations.
\end{abstract}


\section{Introduction}
\label{sec:Intro}
Magnetars, as first discussed by \cite{dt92a}, are
considered to be slowly rotating neutron stars with spin periods
of 5-12 seconds and a rapid spin-down. They exhibit extremely
strong magnetic fields, typically $ > 10^{14} $\,G, the decay of which
is believed to create the observable high X-ray and gamma-ray
luminosities, often visible in bursts.

The model of magnetars tends to fit two previously described
distinct classes of objects, the soft gamma-ray repeaters (SGRs) and
the Anomalous X-ray pulsars (AXPs). Common properties of the members of 
both classes are their long
rotation periods \citep[AXPs; ][]{ms95}, \citep[SGRs; ][]{ksh+99} 
and the bursting nature of their emission \citep{kgw+03,gkw02}.
 They are usually radio quiet. However, the pulsed radio emission
that has been detected from \axp \citep{crh+06} and AXP 1E1547.0$-$5408 
\citep{crj+08} indicates that this is not always the case.

The AXP XTE J1810$-$197 was discovered by \cite{ims+04} in Rossi
X-ray Timing Explorer (RXTE) data of the source SGR~1806$-$20, taken in
January 2003. With a period of 5.54\,seconds and a period derivative
$\sim 1.15\times10^{-11}$\,ss$^{-1}$, a magnetic field of
$\sim 2.6\times10^{14}$\,Gauss is implied.  Although the previous
observations clearly identified it as an AXP, the extreme variation in
the X-ray flux also classified it as the first transient AXP.

The detection of a radio source coincident with the position of \axp by
\cite{hgb+05} first raised the possibility that this was the first radio
emitting magnetar. This possibility was confirmed with the detection of
strong, narrow and highly variable radio pulses, with the same pulse period as
determined at high-energies, by \cite{crh+06}. Further observations have shown
that the emission is $\sim$80-95\,\% polarised, mostly linear, but with a
significant degree of circular polarisation at all observed frequencies
\citep{ksj+07, crj+07}.  One of the most remarkable features of the radio
emission from this source is its flat radio spectrum.  Radio pulsars are
normally characterised by the steepness of their spectrum \citep{mkkw00a,
ljk+08}, which makes their detection at frequencies above 30\,GHz very
difficult. However, soon after the discovery of its radio emission, it became
clear that \axp had a flat radio spectrum $S \propto \nu^{-0.5}$ \citep{crh+06,
crp+07} and for a time became the brightest neutron star at frequencies
greater than about 40\,GHz. Strong variations in the pulsed intensity and the profile
phase were visible in the initial observations \citep{ccr+07}, although it also
became evident that the average pulsed flux density was decreasing
with time.

In this paper we report on the results of simultaneous and
quasi-simultaneous multi-frequency observations conducted at the radio
frequencies of 1.41, 4.90 and 8.35\,GHz during July 2006 and 2.64, 4.85,
8.35, 14.6 and 32\,GHz from December 2006 to July 2007. We present the
radio spectrum of the total radio flux density of \axp and of its
individual profile components and the results of power-law fits to this
spectrum. We also consider the time variability of the fitted spectral index
in light of the modulation of individual pulses
and the intra-day flux density variability. A full discussion of the
individual pulse properties will follow in a separate paper.

\section{Observations}

The simultaneous observations were made using the 100-m radiotelescope
of the Max-Planck Institute for Radioastronomy (MPIfR) at Effelsberg,
Germany, the 76-m Lovell radiotelescope at Jodrell Bank observatory of
the University of Manchester, UK and the 94-m equivalent Westerbork
Synthesis Radio Telescope (WSRT) in the Netherlands. The
quasi-simultaneous observations were made with the Effelsberg
radiotelescope. In total there were 8 simultaneous multi-telescope
multi-frequency sessions during July 2006 and 10 quasi-simultaneous
multi-frequency between December 2006 and July 2007. For the latter
sessions the new sub-reflector of the Effelsberg telescope was
used. It was installed in October 2006, improving the sensitivity and
resulting in flatter gain curves but also allowing fast receiver
changes between secondary and primary focus. The integration time for
every session depended on the observing frequency and the
observational circumstances. In general, the time needed for 1.42,
2.64, 4.85 and 4.90\,GHz was around 5-15\,min, for 8.35\,GHz around
20\,min and for 14.6 and 32\,GHz around 25-40\,min. Details of the
observing sessions are summarised in Tables ~\ref{tab:obs1} and
~\ref{tab:obs2}.

\subsection{Calibration Procedures}

The observing and calibration procedures for Effelsberg data were the
same for the simultaneous and quasi-simultaneous session.  We used the
2.64\,GHz, 4.85\,GHz, 8.35\,GHz, 14.6\,GHz and 32\,GHz cooled HEMT
receivers that are installed in the secondary focus
\citep{khk+01}. In order to calibrate the flux density of a pulsar at
the Effelsberg radio telescope reliably, a noise diode signal is injected into
the wavefront following the horn synchronously with the pulse period
in the beginning of each measurement (Figure~\ref{prof}).  The power
output of the diode is compared with the power that is received from
the pulsar. The signal of the noise diode itself is frequently
calibrated and monitored by observing known reference sources during
regular pointing observations. The sources that were used for
calibration were 3C273, 3C274, 3C286 and NGC7027. They were observed
at the beginning and the end of each session, in combination with
checks on pointing and focus stability. The whole procedure of
Effelsberg calibration is described further in \cite{ang07}. We also
monitored the quality of our observations by taking data for well
known pulsars such as PSR B1929+10 which is strong enough to be
detected at all frequencies up to 43 GHz. The observed properties of
this pulsar, extensively also studied at Effelsberg, were compared to
archival data to confirm that the system was functioning correctly at
all frequencies. The observing and calibration procedures for the
other telescopes participating in our simultaneous observations
between May and July 2006 are described for each telescope in detail
in \cite{ksj+07}.

\begin{table}
\caption{\label{tab:obs1} Summary of simultaneous observing sessions in July 2006.}
\begin{tabular}{cclcc}
\hline
{Date} & {Session} & {Telescope} &{Frequency}   & {BW} \\
             &                   &              & {(GHz)}          & {(MHz)} \\
\noalign{\smallskip}
\hline
\noalign{\smallskip}
31/05/06 & 1 & Lovell &   1.42& 32\\
             &    & WSRT & 4.90& 160\\
            &     & Effelsberg     &8.35 & 1000\\
\noalign{\smallskip}
10/07/06 & 2 & Lovell & 1.42& 32\\
            &     & Effelsberg     &8.35 & 1000\\
\noalign{\smallskip}
17/07/06 & 3 & Lovell & 1.42& 32\\
            &    & WSRT & 4.90& 160\\
            &     & Effelsberg     &8.35 & 1000\\
\noalign{\smallskip}
21/07/06 & 4 & Lovell & 1.42& 32\\
            &     & WSRT  &4.90 & 160\\
\noalign{\smallskip}
22/07/06 & 5 & Lovell & 1.42& 32  \\
             &    & WSRT & 4.90& 160\\
            &     &Effelsberg  &8.35 & 1000\\
           &     &Effelsberg  &14.60 & 2000\\
\noalign{\smallskip}
23/07/06 & 6 & Lovell & 1.42& 32\\
            &     & WSRT &4.90 & 160\\
\noalign{\smallskip}
26/07/06 & 7 & Lovell       & 1.42 & 32  \\
              &    & Effelsberg  & 4.85 & 500 \\
              &     &    WSRT  & 4.90 & 160 \\
             &     &  Effelsberg & 8.35 & 1000 \\
             &    &   Effelsberg & 14.60& 2000 \\
\noalign{\smallskip}
28/07/06 &8 &  Lovell & 1.42 & 32\\
             &     & Effelsberg   & 4.85 &  500 \\
             &     & WSRT      & 4.90 &  160 \\
             &     & Effelsberg & 8.35 & 1000 \\
             &    & Effelsberg  & 14.60& 2000 \\
\noalign{\smallskip}
\hline
\end{tabular}
\end{table}
\begin{table}
\caption{\label{tab:obs2} Summary of quasi-simultaneous observing sessions from December 2006 to July 2007.}
\begin{tabular}{cclcc}
\hline
{Date} & {Session} & {Telescope} &{Frequency}   & {BW} \\
             &                   &              & {(GHz)}          & {(MHz)} \\
\noalign{\smallskip}
\hline
\noalign{\smallskip}
09/12/06 & 1 &  Effelsberg& 2.64 & 100 \\
              &    &           & 4.85 & 500  \\
             &     &           & 8.35 & 1000 \\
            &     &           &14.60 & 2000\\
\noalign{\smallskip}
26/12/06 & 2 &  Effelsberg& 2.64 & 100 \\
              &    &           & 4.85 & 500  \\
             &     &           & 8.35 & 1000 \\
            &     &           &14.60 & 2000\\
\noalign{\smallskip}
04/02/07 & 3 &  Effelsberg& 2.64 & 100 \\
              &    &            & 4.85 & 500  \\
             &     &           & 8.35 & 1000 \\
            &     &            &14.60 & 2000\\
\noalign{\smallskip}
06/02/07 & 4 & Effelsberg  & 4.85 & 500 \\
             &     &              & 32.00 & 2000 \\
\noalign{\smallskip}
12/02/07 & 5 & Effelsberg  & 4.85 & 500 \\
             &     &              & 14.60 & 2000 \\
\noalign{\smallskip}
17/02/07 & 6 &  Effelsberg& 4.85 & 500\\
              &    &            & 8.35 & 1000 \\
            &     &          &14.60 & 2000\\
             &     &            & 32.00 & 2000 \\
\noalign{\smallskip}
18/02/07 & 7 &  Effeslberg& 4.85 & 500\\
              &    &          & 8.35 & 1000 \\
            &     &          &14.60 & 2000\\
             &     &           & 32.00 & 2000 \\
\noalign{\smallskip}
26/03/07 & 8 &  Effelsberg& 4.85 & 500\\
            &     &                &14.60 & 2000\\
             &     &             & 32.00 & 2000 \\
\noalign{\smallskip}
05/05/07 & 9 &  Effelsberg& 4.85 & 500\\
            &     &                &8.35 & 1000\\
             &     &             & 32.00 & 2000 \\
\noalign{\smallskip}
06/07/07 & 10 &  Effelsberg& 4.85 & 500\\
            &     &              &8.35 & 1000\\
             &     &           & 14.60 & 2000 \\
\noalign{\smallskip}
\hline
\end{tabular}
\end{table}

\section{Data Analysis \& Results}

\begin{figure}
\mbox{\includegraphics{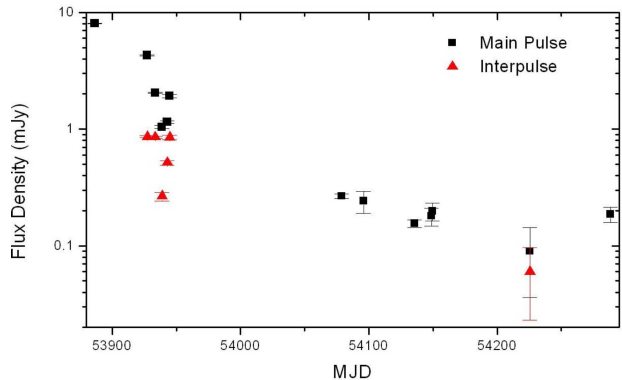}}
\caption{The average flux density for each observing session as
measured at 8.35 GHz for the main pulse and the interpulse (when detected).}
\label{avflux}
\end{figure}

\begin{figure}
\mbox{\includegraphics{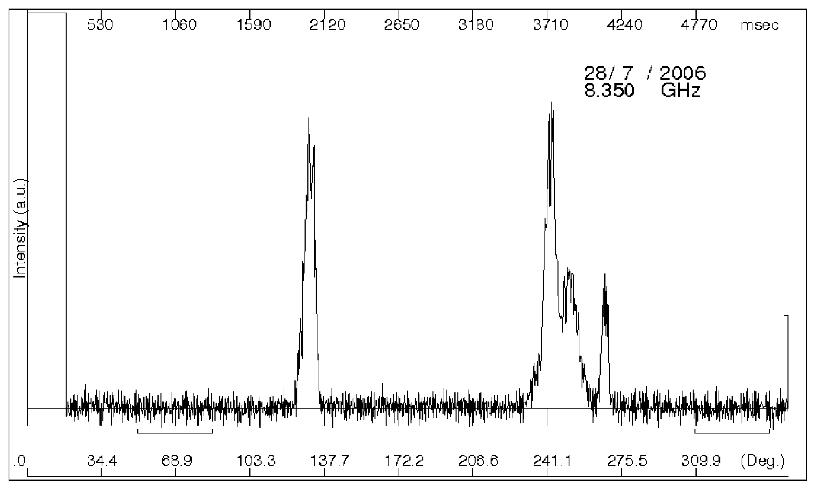}}  
\mbox{\includegraphics{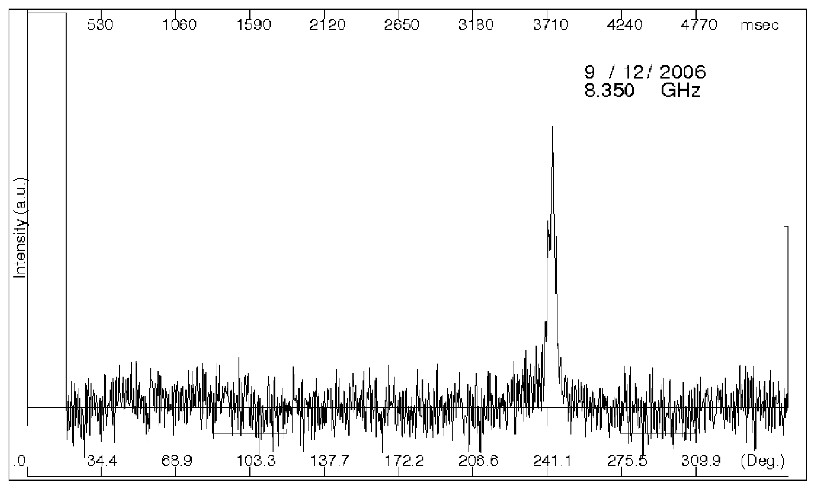}} 
\mbox{\includegraphics{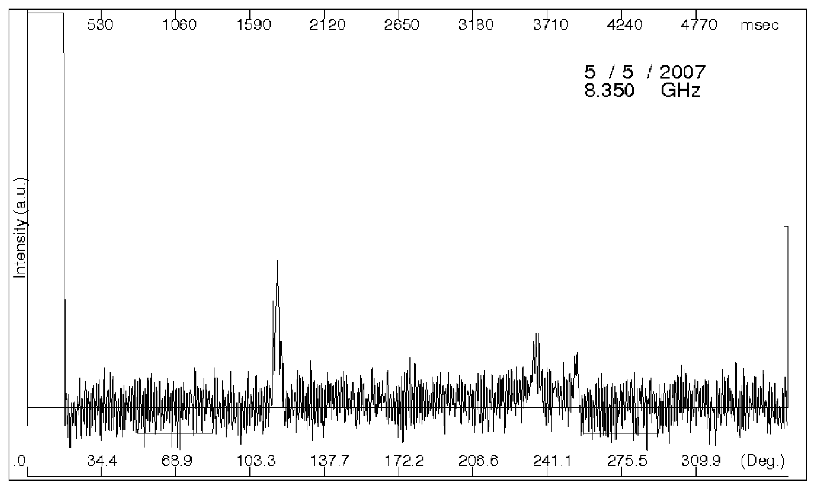}}
\caption{Phase aligned integrated profiles of \axp in 2006 July (top), 2006
  December (middle) and 2007 May (bottom). The source becomes significantly
  weaker while the IP eventually disappears. Only in 2007 May, both the MP and
  the IP are visible again. Note that each shown profile contains the signal
  of the calibrating noise diode which was switched on synchronously with the
  pulse period for the first 50 phase bins.}
\label{prof}
\end{figure}

Until August 2006, the integrated pulse profile of \axp consisted of
two major well-separated features which we discussed in some detail
already in \cite{ksj+07}. Following the same convention, we refer to
them as the {\em main pulse} (MP, right feature in Figure~\ref{prof}a)
and {\em interpulse} (IP, left feature in Figure~\ref{prof}a). The MP
is the more complex and wider of the two, with a varying width of
about 0.7 seconds (or 45 deg in pulse longitude) while the IP is much
narrower with a width of typically 0.2 seconds (or 12 deg longitude).
In some cases, three or more distinct sub-components were visible in
the MP.  The simpler IP was not always visible and was only strongly
detected during parts of our observations, as indicated in
Figure~\ref{avflux} where we present the daily average flux densities
of the two components as measured at 8.4 GHz as a function of
time. The identification of the remaining visible pulse feature (see
e.g.~Figure~\ref{prof}b) with the MP is possible due to timing
information obtained from regular monitoring observations with the
Lovell telescope at the Jodrell Bank observatory (Lyne et al. in
preparation). After the disappearance of the IP in summer 2006, both
MP and IP were visible again simultaneously only during one short
session in May 2007. The IP remains undetected since.  

Overall, the flux density of the source has significantly decreased
since its first detection at radio frequencies (see Figure~\ref{avflux}),
which is consistent with the earlier findings by \cite{ccr+07}.  Due to differences in
the frequency range of the simultaneous and quasi-simultaneous
observations we discuss further results from these campaigns
separately.

\subsection{Simultaneous Observations}


\begin{figure}
\mbox{\includegraphics{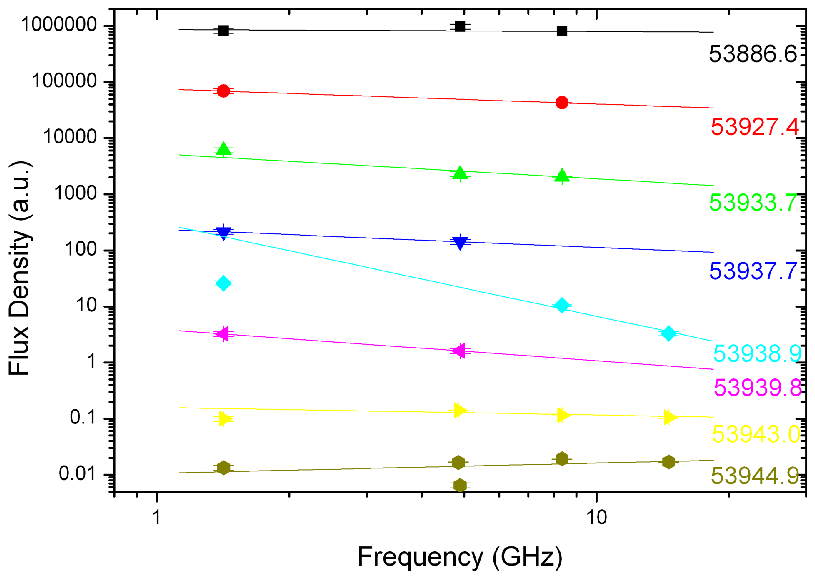}}
\mbox{\includegraphics{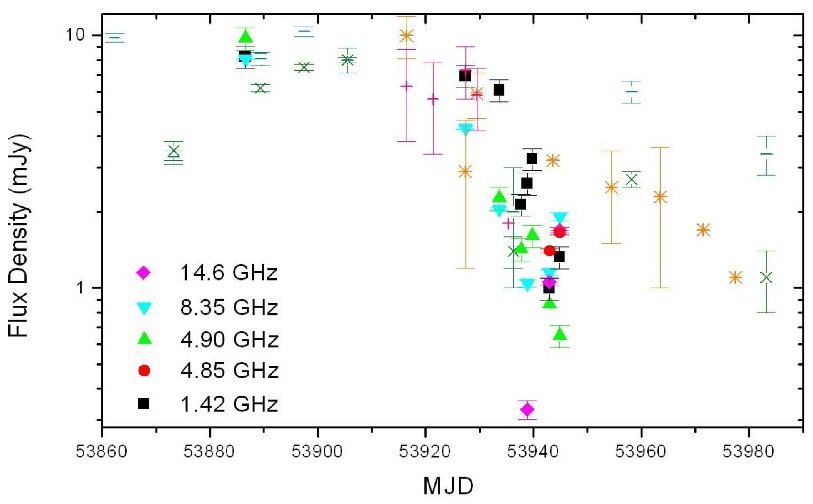}}
\mbox{\includegraphics{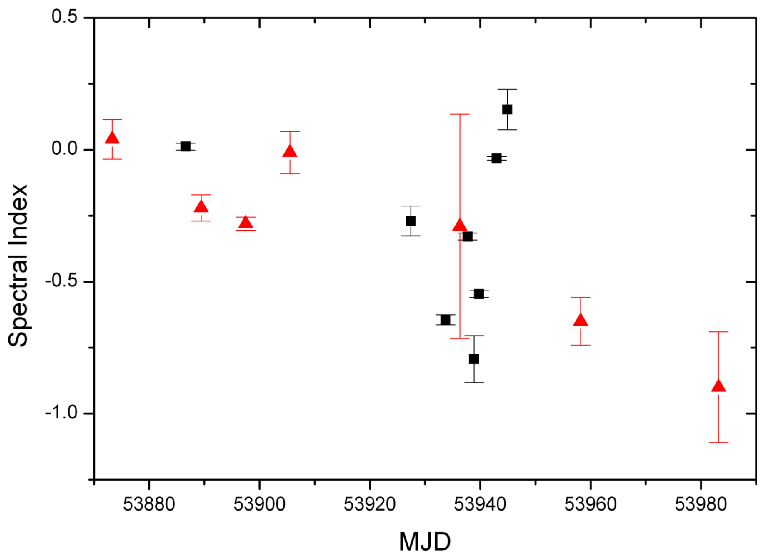}}
\caption{(Top) Flux density spectrum of the MP as determined during our simultaneous
  observations. For plotting purposes the flux densities of each day were
  multiplied by a different factor to distinguish the dates more
  clearly. (Middle) Flux density and its variability as measured at various
  frequencies as a function of time. We also added data measured by
  Hotan et al.~(2007) at 1.4 and 4.8\,GHz (shown as crosses and
  stars, respectively) and Camilo et al.~(2007c) at 1.42 and 4.8\,GHz (shown as bars and
  exes, respectively). (Bottom) Derived spectral indices of the MP as a
  function of time. The average value is determined $\alpha = -0.31
  \pm0.06$. In comparison, we show spectral indices derived by Camilo et al.~(2007c)
  from VLA measurements as triangles.}
\label{MP}
\end{figure}

\begin{figure}
\mbox{\includegraphics{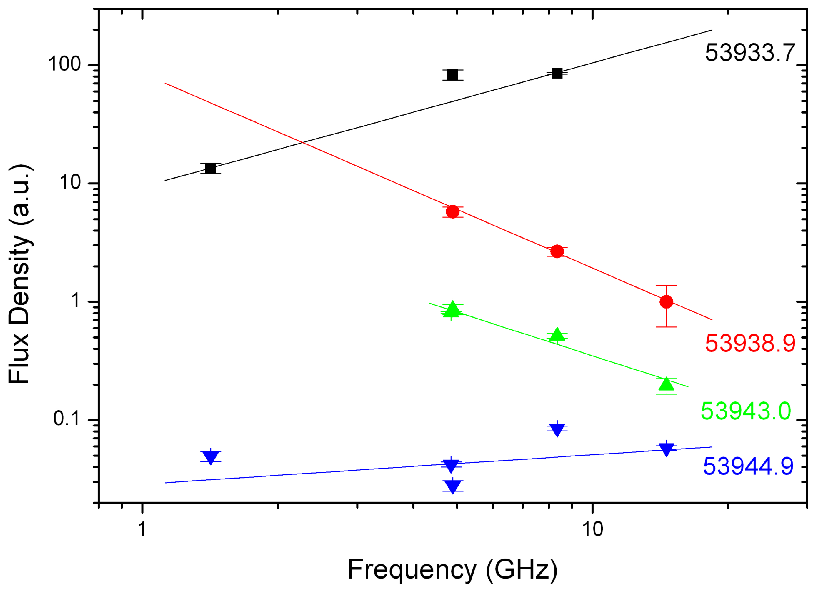}}
\mbox{\includegraphics{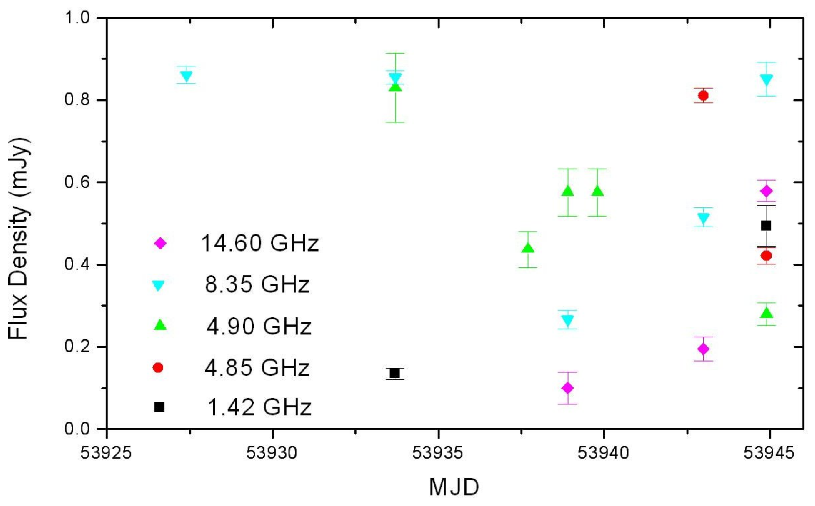}}
\mbox{\includegraphics{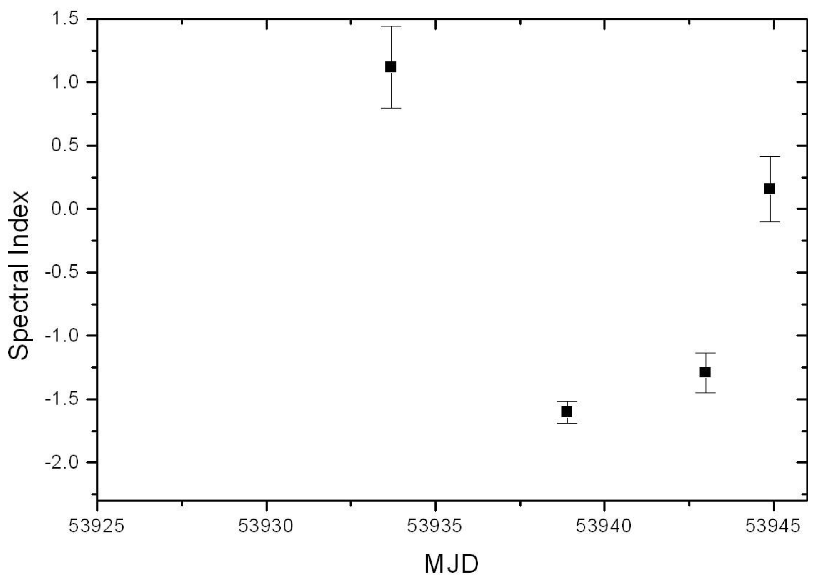}}
\caption{(Top) Flux density spectrum of the IP as determined during our simultaneous
  observations. For plotting purposes the flux densities of each day were
  multiplied by a different factor to distinguish the dates more
  clearly. (Middle) Flux density and its variability as measured at various
  frequencies as a function of time.
  (Bottom) Derived spectral indices of the IP as a
  function oft time. The average value is determined $\alpha = -0.41 \pm
  0.21$. }
\label{IP}
\end{figure}

The flux densities measured during the simultaneous observations and
the spectral indices derived from power law fits, $S\propto
\nu^\alpha$) to these data, are listed in Table~\ref{tab:resm} and are
shown in Figures \ref{MP} and \ref{IP}. For each observation we summed
all individual pulses to obtain an integrated total power profile that
was flux calibrated. The nominal conservative uncertainties for our
flux density measurements are about 10\% (see also Section 3.2). In
addition to our flux densities, we also utilize published flux
densities measured at 1.42 and 4.8\,GHz with the VLA \citep{crp+07}
and at 1.4 and 4.8\,GHz at the Mount Pleasant observatory
\citep{hldd07} (see Fig.~\ref{MP}).  Our flux densities observed for
the MP at similar epochs for these frequencies are in good agreement
with these values.  This fact indicates that the effect of
interstellar scintillation may be small due to the large observing
bandwidths, but we will discuss the possible impact of the
interstellar medium on our results in more detail later.  Overall, a
large decrease in the signal strength with time is observed. At the
same time, the spectrum is generally flat with an average spectral
index of $\alpha = -0.31 \pm 0.06$ for the MP and $\alpha = -0.41 \pm
0.21$ for the IP (see Fig.~\ref{MP} and ~\ref{IP}), consistent with results by
\cite{crh+06, crp+07}.  However, around MJD 53938 (2007 July 22) we
observe a strong time variability, with the spectrum being initially
steep and then flattening day by day for both MP and IP.

\begin{table}
\caption{\label{tab:resm} Summary of flux densities and spectral indices for the main pulse.}
\begin{tabular}{ccccc}
\hline
{Date} & {Frequency}   & ${S_{m}}$ & $ {\alpha} $ \\
             & {(GHz)}          & {(mJy)}  &      \\
\noalign{\smallskip}
\hline\\
\noalign{\smallskip}
31/05/06 & 1.42 & $8.23	\pm 0.82 $& $ +0.01 \pm 0.11$ \\
             &  4.90& $9.73 \pm	0.97$ & \\
            & 8.35 & $8.03	\pm 0.01$ &              \\
\noalign{\smallskip}
10/07/06 & 1.42 & $6.90 \pm 	0.69$  	 & $-0.27 \pm	0.06$ \\
            & 8.35 & $4.28 \pm	0.03$	  &              \\
\noalign{\smallskip}
17/07/06 & 1.42 & $6.06	\pm 0.61$ 	 & $-0.64	\pm 0.14$ \\
             &  4.90 & $2.27 \pm	0.23$ & \\
            & 8.35 & $2.05 \pm 0.03$&              \\
\noalign{\smallskip}
21/07/06 & 1.42 &  $2.14 \pm 0.21$ & $-0.33 \pm 0.11$  \\
            & 4.90 &  $1.42 \pm 0.14$  &                        \\
\noalign{\smallskip}
22/07/06 & 1.42 &  $2.59 \pm 0.26$& $ -0.79 \pm 0.30$ \\
             & 8.35 & $1.04 \pm 0.03$    &                 \\
            & 14.60 & $0.33 \pm  0.03$  &                        \\
\noalign{\smallskip}
23/07/06 & 1.42 & $3.24 \pm 0.32$ & $-0.55 \pm 0.11$  \\
            & 4.90 & $1.61 \pm 0.16$ &              \\
\noalign{\smallskip}
26/07/06 & 1.42 & $0.99 \pm	0.10$ &$-0.03 \pm 0.11$ \\
             & 4.85 &$1.40 \pm 0.03    $ &                \\
             &4.90 &$0.86 \pm 0.09$&                     \\
             &  8.35 &$ 1.15 \pm 0.03  $&                      \\
            &  14.60 &$ 1.06 \pm 0.04  $&                        \\
\noalign{\smallskip}
28/07/06 & 1.42 &$1.32 \pm 0.13   $&$0.15 \pm	0.28$ \\
              & 4.85 &$1.66 \pm 0.03     $&                \\
              & 4.90 &$0.65 \pm 0.06     $&                \\
             &  8.35 &$1.92 \pm 0.07     $&                         \\
          & 14.60 &$1.70 \pm 0.04        $ &                          \\
\noalign{\smallskip}
\hline
\end{tabular}
\end{table}
\begin{table}
\caption{\label{tab:resi} Summary of flux densities and spectral indices for the interpulse.}
\begin{tabular}{ccccc}
\hline
{Date} & {Frequency}   & ${S_{m}}$ & $ {\alpha} $ \\
             & {(GHz)}          & {(mJy)}  &      \\
\noalign{\smallskip}
\hline\\
\noalign{\smallskip}
10/07/06 & 8.35 & $0.86	\pm 0.02 $  	 &   -   \\
\noalign{\smallskip}
17/07/06 & 1.42 & $0.13 \pm	0.01$  	 & $ +1.12 \pm	0.33 $ \\
             & 4.90 & $0.83 \pm 0.08$  	 &          \\
            & 8.35 & $0.86 \pm 0.02$    &              \\
\noalign{\smallskip}
21/07/06 & 4.90 &  $0.44 \pm 0.04$ &       -     \\
\noalign{\smallskip}
22/07/06 & 4.90 &  $0.58	 \pm 0.06$ &  $-1.61 \pm	0.08$  \\
             & 8.35 & $ 0.27 \pm 0.02$&           \\
            & 14.60 &$ 0.10 \pm 0.04$ &                        \\
\noalign{\smallskip}
23/07/06 & 4.90 & $0.58 \pm 0.06$ &      -        \\
\noalign{\smallskip}
26/07/06 & 4.85 & $0.81 \pm 0.02$ &$-1.29 \pm	0.26$ \\
             &  8.35 & $0.51 \pm  0.02$ &                      \\
            &  14.60 & $ 0.19 \pm 0.03 $  &                        \\
\noalign{\smallskip}
28/07/06 & 1.42 &$0.49 \pm 0.05$   &$0.15 \pm	0.28$ \\
              & 4.85 &$0.42 \pm 0.02$    &                      \\
              & 4.90 &$0.28 \pm 0.03$     &                       \\
             &  8.35 &$0.85 \pm 0.04$       &                         \\
          & 14.60 &$0.58 \pm 0.03$          &                          \\
\noalign{\smallskip}
\hline
\end{tabular}
\end{table}

\subsection{Quasi-Simultaneous Observations}

\begin{figure}
\mbox{\includegraphics{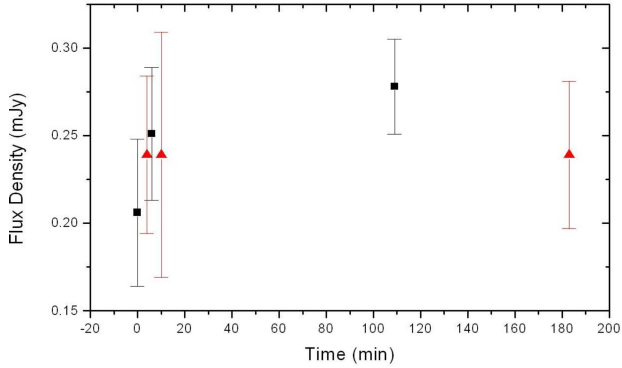}}
\caption{ Flux density measured at 4.85\,GHz during an observing session for
  two consecutive days (square and up triangle points). The first
  two measurements for each day correspond to a 5 minutes integration time 
  while the third measurement corresponds to 15 minutes.}
\label{quasitime}
\end{figure}

All quasi-simultaneous observations were performed with Effelsberg alone.
The short time needed to switch between the different receivers (i.e.~$\sim
30-60$s between secondary focus receivers, and $\sim 2-4$min between secondary
and primary receivers) made it possible to observe at a wide range of
frequencies in a single observing session. The longest multi-wavelength
observing session (without the calibration scans) of \axp was about four
hours, during which frequencies were cycled through repeatedly in order to
detect any short-term variability. Using these data, we not only studied the
long term variability of \axp, but also the medium-term intra-day flux density
fluctuations as well as the modulation of individual pulses during each of
these sessions. An example of such measurements is shown in
Fig~\ref{quasitime} where we present the flux density as measured repeatedly
at 4.85\,GHz over two consecutive days. The observed variation is
consistent with a constant flux density over this period.  The results of all
our quasi-simultaneous observations are summarized in Table~\ref{tab:res}.

\begin{figure}
\mbox{\includegraphics{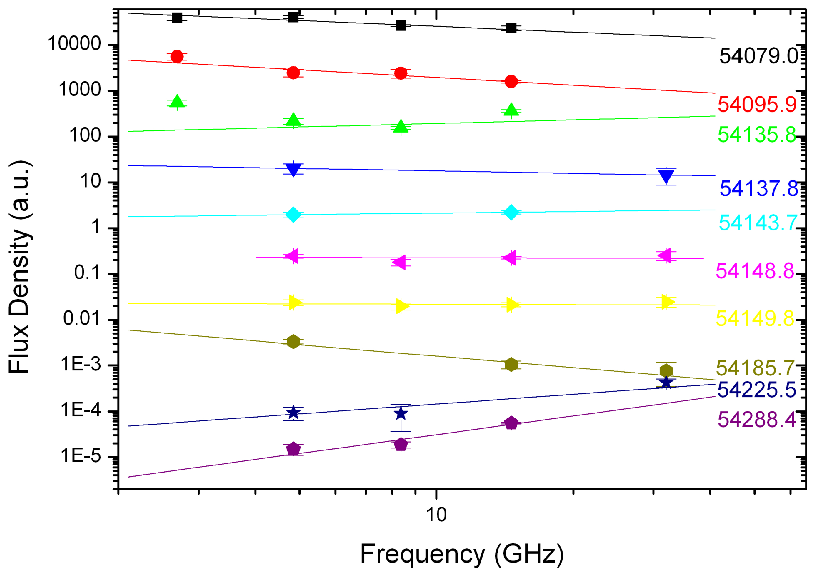}}
\mbox{\includegraphics{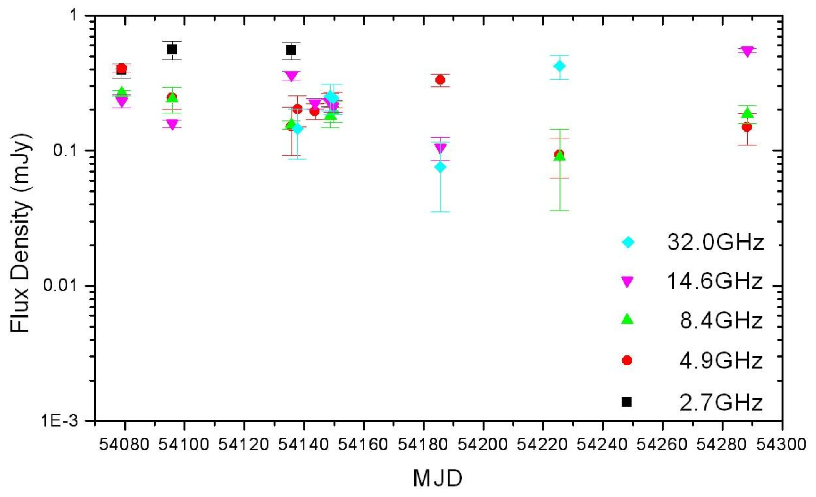}}
\mbox{\includegraphics{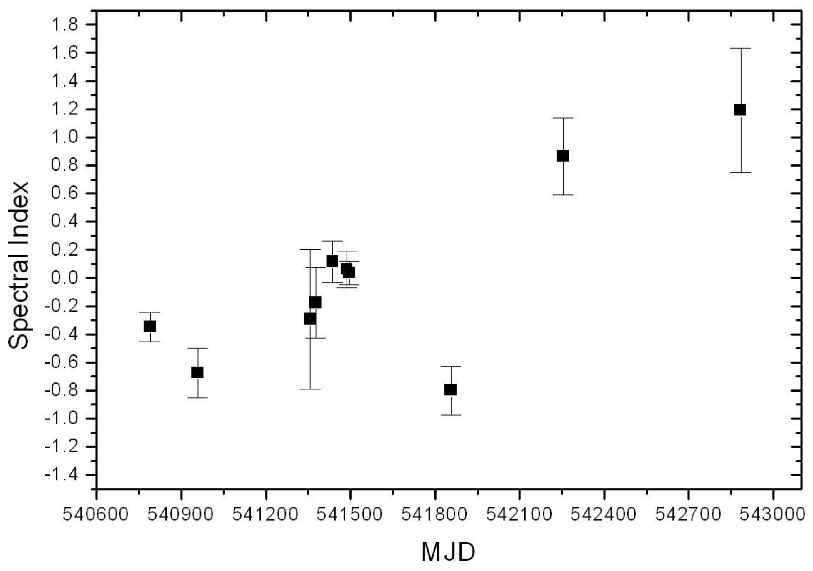}}
\caption{(Top) Flux density spectrum of the MP as determined during our
  quasi-simultaneous observations. For plotting purposes the flux densities
  for each day were multiplied by a different factor to distinguish the dates
  more clearly. (Middle) Flux density and its variability as measured at
  various frequencies as a function of time.  (Bottom) Derived spectral
  indices of the MP as a function oft time. The nominal average value is
  determined as $\alpha = -0.003 \pm 0.089$. }
\label{spctrm}
\end{figure}

The average flux density values were then used to determine the
spectra and their indices. This was done for each session and the
results are shown in Figure \ref{spctrm}. The results are consistent
with and extend those of the simultaneous measurements.  We find again
that the spectrum is generally flat with a mean spectral index of
$\alpha = 0.00 \pm 0.09$ but it is also variable on a day-to-day basis
with significant variations around its mean value. This is reflected
also by the variation of the flux densities measured at 8.35\,GHz and
above which appear to sometimes follow a common pattern that is
anti-correlated with changes seen below that frequency. There is an
indication of a general increase of $\alpha$ over the 210 days covered
by our observations, starting with initial values around $\alpha \sim
-0.2$ and reaching $\alpha \sim 1.2$ near the end of the observing
period.

\begin{table}
\caption{\label{tab:res} Summary of flux densities and spectral indices
for the MP measured during our quasi-simultaneous observing sessions.}
\begin{tabular}{ccccc}
\hline
{Date} & {Frequency}   & ${S_{m}}$ & $ {\alpha} $ \\
             & {(GHz)}          & {(mJy)}  &      \\
\noalign{\smallskip}
\hline\\
\noalign{\smallskip}
09/12/06 & 2.64 & $ 0.39	\pm 0.05$ & $-0.35 \pm 0.11$ \\
              & 4.85 & $ 0.40	 \pm 0.03$ &                       \\
             &  8.35 & $ 0.27	\pm 0.01$ &                         \\
            & 14.60 & $ 0.23	 \pm 0.03$ &                        \\
\noalign{\smallskip}
26/12/06 & 2.64 & $0.56 \pm 0.09 $&$ -0.67 \pm 0.17$ \\
              & 4.85 & $0.25	\pm 0.05 $&                        \\
             &  8.35 & $0.24 \pm	 0.05$ &                      \\
            &  14.60 & $0.16 \pm  0.01$ &                        \\
\noalign{\smallskip}
04/02/07 & 2.64 &$ 0.55	\pm 0.08 $&$ -0.24 \pm 0.49$ \\
              & 4.85 &$ 0.15\pm 0.05$ &                        \\
             &  8.35 &$ 0.15	\pm 0.01$ &                         \\
            & 14.60 &$ 0.36 \pm 0.03$ &                          \\
\noalign{\smallskip}
06/02/07 & 4.85 & $0.20	\pm 0.05$&$ -0.18 \pm 0.25$\\
             & 32.00 &$ 0.14	\pm 0.06 $&                  \\
\noalign{\smallskip}
12/02/07 & 4.85 &$ 0.20 \pm 0.03 $&$ 0.11 \pm 0.15$\\
             &14.60 &$ 0.22	\pm 0.02$ &                      \\
\noalign{\smallskip}
17/02/07 & 4.85 & $0.24  \pm  0.02$ &$ +0.06	 \pm  0.13$\\
              & 8.35 & $0.18 \pm 	 0.03$&                       \\
              &14.60 & $0.23 \pm  0.01 $&                       \\
             &  32.00 & $0.25	\pm  0.05$&                      \\
\noalign{\smallskip}
18/02/07 & 4.85 & $0.24  \pm  0.03 $& $+0.03  \pm  0.08$ \\
              & 8.35 &$ 0.20  \pm  	0.03$ &                      \\
             &14.60 &$ 0.21  \pm  0.02 $&                       \\
             &  32.00 & $0.25  \pm  0.06 $&                      \\

\noalign{\smallskip}
26/03/07 & 4.85 & $0.33  \pm  	0.03$ &$ -0.80  \pm 	0.17$\\
             & 14.60 &$ 0.10  \pm  	0.02$ &                           \\
             & 32.00 &$ 0.08  \pm  0.04 $&                         \\
\noalign{\smallskip}
05/05/07 & 4.85 & $0.09  \pm  	0.03$ &$ +0.86  \pm 	0.27$\\
             & 14.60 &$ 0.09  \pm  	0.05$ &                           \\
             & 32.00 &$ 0.42  \pm  0.08 $&                         \\
\noalign{\smallskip}
06/07/07 & 4.85 & $0.15  \pm  	0.04$ &$ +1.20  \pm 	0.44$\\
             & 8.35 &$ 0.19  \pm  	0.03$ &                           \\
             & 14.60 &$ 0.55  \pm  	0.02$ &                           \\
\noalign{\smallskip}
\hline
\\
\end{tabular}
\end{table}

\subsection{Modulation Indices}

In order to characterise the observed flux density variations and to obtain
reliable estimates for the measurement uncertainties, we also studied the
modulation of the {\em single pulse} flux densities (see e.g.~\cite{kkg+03} and
\cite{skmj07}).
We can use these results also to estimate the impact of the
interstellar medium on our measurements, following similar studies such as that
conducted for pulsars by \cite{mss+96} at very similar frequencies,
i.e.~4.75\,GHz and 10.55\,GHz.

We concentrate on the flux densities measured at 4.85\,GHz and 14.6\,GHz
during our quasi-simultaneous sessions. The flux densities at these
frequencies were measured in every session, often at the beginning and the end
of a session, providing a densely sampled data set that gives a good
representation of the behaviour between widely spaced frequencies.

As a comparison, we also studied data for our reference source PSR B1929+10 to
rule out systematic effects, impact of weather, or instabilities in the
receivers chain. We find, as expected, only minor variations in the flux
density of this well known pulsar at high frequencies, in line with the 
findings of \cite{mss+96}.

\begin{figure}
\mbox{\includegraphics{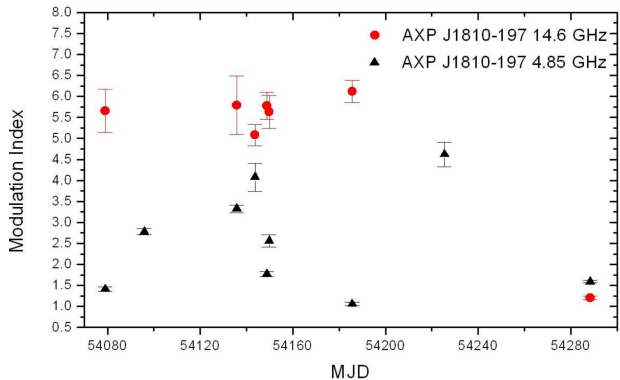}}
\caption{Variation of the pulse-to-pulse modulation index of \axp at 4.85\,GHz
  (squares) and 14.6\,GHz (triangle). At 14.6\,GHz the indices were computed
  for time series covering 25-30 min while the 4.85\,GHz data cover 10-30
  mins.  }
\label{Mod}
\end{figure}

For each session, we calculated the pulse-to-pulse modulation index $m$
according to $m^2=\frac{\langle(S-<S>)^2\rangle}{<S>^2}=\sigma_S^2/<S>^2$
(e.g.~Kramer et al.~2003)\nocite{kkg+03}, where $S$ is the measured flux
density, $\langle S \rangle$ its mean value and $\sigma_S$ its standard
deviation. We present the results in Figure~\ref{Mod}.  As already expected
from our discussion in \cite{ksj+07} and consistent with early observations of
\axp by \cite{crh+06}, the pulses are very highly modulated and variable for
both frequencies. For the densely covered period of time around epoch
MJD 54150 (04 February-26 March 2007) the data occasionally suggest a behaviour that is
anti-correlated between 4.85 GHz and 14.6 GHz. Overall, the higher frequency
shows greater modulation, but it is possible that weaker single pulses are more
difficult to detect at these frequencies. Despite this large modulation of the
single pulses, averaging over sufficient time as done during all our
observations essentially removes the variation for a given
measurement value, leading to relative uncertainties in $S$ of the order
$m/\sqrt{n}$ where $n$ is the number of averaged pulses. This is consistent
with our estimated flux density errors and confirmed by the repeatability of
our flux density measurements over consecutive days (see Fig~\ref{quasitime}).
We therefore expect the impact of slow-varying effects caused by refractive
scintillation to be more important. We will discuss this in the following.

\section{Discussion}

Our monitoring observations of \axp and its flux density spectrum over the
course of more than one year appears to confirm the conclusion derived already
in all previous studies of it,
that \axp behaves unlike most or even any known radio pulsar. Several observed
properties support this fact. In our range of studied frequencies, the source
exhibits a generally flat flux density spectrum, with values of spectral
index consistent with those measured by \cite{crh+06, crp+07}. These flat
spectra allowed us to observe the source up to very high frequencies, at least
when compared to radio frequencies typically used for ordinary pulsar
observations. An example is our detection of the magnetar at 43 GHz in May
2007 with the Effelsberg radiotelescope. This is only the fifth neutron star
detected at $\lambda7$mm \citep{kjdw97} and the only one where single pulses
could be observed.  This is consistent with observations at the IRAM 30\,m
telescope at 88 and 144\,GHz by \cite{crp+07}, representing the highest radio
frequency at which a neutron star has been detected. For normal pulsars, we
often see breaks in their power law spectra at high frequencies
\citep{mkkw00a, kgk07} or even occassionally a turn-up at mm-wavelengths
\citep{kjdw97}. Here, in our range of frequencies, the spectrum of \axp is
well described by a single, flat power law, of the form $S_{\nu} \propto \nu^\alpha$
with $\alpha = 0.0 \pm 0.5$.

In those cases where both the MP and IP were detected, their spectral indices
were found to be differing, with the IP showing a significantly greater
variation. Even though the spectra appear to change in the same manner from
day to day, becoming steeper or flatter together, the IP exhibited many more
extreme positive and negative steep values. This is somewhat surprising since
the MP spectral index reflects emission from a wider range of emission
components, in comparison to an IP that is always observed as a rather simple
emission feature. The variation of the spectrum is roughly
  consistent with the model by Thompson (2008)\nocite{tom08} where timescales
  of $\sim0.1$ day may be explained by current-driven instabilities on the
  closed magnetic field lines. The observation that it affects the IP
  and MP in a different way is interesting and may support the idea of a
  rather twisted magnetosphere. Although the spectrum is generally flat, 
the spectral index shows significant variation, with a slight 
trend of becoming more positive with time. 

While we have shown that the effect of the significant pulse-to-pulse
modulation can be removed by averaging a sufficient number of pulses,
another well known cause for flux density variations, and hence
possible spectral index variations, is interstellar scintillations
(ISS).  Based on studies of the ISS at 4.75 GHz and 10.55 GHz by
\cite{mss+96}, the relatively large dispersion measure of
DM~=~178\,pc\,cm$^{-3}$ and the estimated distance of 3.3\,kpc,
suggests a critical scintillation frequency $f_c$ of $\sim$14\,GHz, well
within our observing band at 14.6\,GHz.  At this frequency, we may
therefore expect a large variation of the measured flux density,
whereas above the critical frequency we expect low flux density
variations due to weak scintillation. Below $f_c$ strong scintillation
will occur, with a branch due to diffractive scintillation and one due
to refractive scintillation.  \cite{lk05} use a different functional
dependence for the transition frequency which yields an $f_c \simeq
60$\,GHz which agrees with the estimates by \cite{crp+07} but is
also smaller than $f_c \simeq 140$\,GHz as derived from the NE2001 electron
density model \citep{cl02}. Despite these differences, the usage of
large integration times and observing bandwidths will effectively
average a number of scintils, resulting in reliable flux density
measurements if bandwidth and integration time are sufficiently
large. Following \cite{lk05}, at 15\,GHz we estimate a diffractive
scintillation bandwidth $\Delta\nu_d =320\,$\,MHz and a diffractive
scintillation timescale of $t_d = 600$\,s which is in good agreement
with the characteristics of features seen in the dynamic spectrum at
the same frequency by \cite{rch+07}. Our observations always average
over several diffractive scintles in frequency and in time, but will
be affected by refractive scintillations at high frequencies. Around
15\,GHz, the modulation index for refractive scintillations is
estimated as $m_r = 0.6$ and the timescale for refractive modulations
turns out to be $t_r = 12.8\,$\,hours. Our observatories could not track
the source for such a long time, so that we account for these possible
variations with an increased quoted flux density error. For typical
integration times of $t_{int}= 40\,$min, our observing set-up should
result in typical errors of our individual flux density measurements
due to refractive and diffractive scintillations of about 20\% at
1.4\,GHz, rising to a maximum of 77\% at 14.6\,GHz and levelling off
at about 62\% at 33\,GHz and above.  Similar estimates for the low-DM
reference source PSR B1929+10 are in good agreement with the
observations. We therefore conclude that the ISS plays a significant
role in the individual flux density measurements, although it cannot
be responsible for the pulse-to-pulse modulation index, and the
described variations of the spectral index for different features of
the profile between individual sessions.

Considering the overall flux density of \axp, we can divide our
observations, spanning more than a year, in four intervals. The first lasts until
July 2006 when the average flux density from the source was above 6 mJy. The
second lasted from July to September 2006 when the average flux density was
above 1 mJy. The third epoch lasted from October 2006 to July 2007 when the
flux density was below 0.5 mJy and, finally the time after July 2007 when the
source became too weak for regular detection. At the same time, we observe a
trend that the flux density variations are larger in the low-flux stages.
This naturally makes the flux density spectrum computation less certain.

Due to the variability of the flux densities, it is difficult to compute an
average spectrum. Especially when we inspect the flux density measurements
from the third epoch, we see many cases in which the high and the low
frequencies vary in a completely different manner for the same day. As we have
seen, some anti-correlation in flux density variations between high and low
frequencies appears present, and given our discussion above, we consider these
variations to be intrinsic to the source and not due to ISS. They can in part
be explained by the assumption that we observe different components of the MP
on different days. However, a similar variation is also observed for the IP,
where always the same emission component is visible. Therefore, we conclude
that in contrast to pulsars, the radio emission of magnetars is not
intrinsically stable. That is consistent with the overall decay of the flux
density in recent months and may indicate that the radio emission is a
transient phenomenon that was triggered by the high-energy outburst. As
the conditions for radio emission may revert back to the pre-outburst stage, it
will be interesting to monitor the source during and after the next outburst.

\section{Summary}

As a result of a coordinated measurement campaign of three telescopes
operating simultaneously at four different frequencies and several daily
Effelsberg measurements up to frequencies of 43\,GHz, we find \axp to be an
unusual pulsating radio source. Its spectral properties and temporal
fluctuations differ remarkably from normal pulsars.

A complex picture of the variability of the radio flux density emerges as a
result of our observations. Significant variability exists on all considered
time scales, from pulse to pulse, day-to-day and over the time of weeks and
months, most of being it due to intrinsic variations (profile changes) and
only some of it affected by scintillations.

Normal pulsars have stable average profiles enabling us to model 
their emission mechanism. This is however not possible with this
source. Also the very flat spectrum and the visibility to high frequencies
makes the source unique among radio emitting neutron stars.

\section*{Acknowledgements}

Kosmas Lazaridis was supported for this research through a stipend from the
International Max Planck Research School (IMPRS) for Astronomy and Astrophysics
at the Universities of Bonn and Cologne. We especially thank
Emmanouil Angelakis for the programming help and for the fruitful
discussions. Maciej Serylak was supported by the EU Framework 6 Marie Curie
Early Stage Training programme under contract number MEST-CT-2005-19669
"ESTRELA".


\end{document}